\documentclass[%
reprint,
superscriptaddress,
 amsmath,amssymb,
aps,
prb,
]{revtex4-1}

\usepackage{graphicx}
\usepackage{epstopdf}
\usepackage{dcolumn}
\usepackage{bm}
\usepackage{hyperref}

\newcommand {\nc} {\newcommand}
\newcommand {\rn} {\renewcommand}

\nc{\bittot}{Bi$_2$Sr$_2$CaCu$_2$O$_{8+\delta}$}
\nc{\capa}{\mathcal{C}_n}
\nc{\caparison}{\mathcal{C}_n(\vec{k}, \omega)}
\nc{\cuotwo}{CuO$_2$}
\nc{\ef}{E_F}
\nc{\ek}{\varepsilon(\kvec)}
\nc{\hightc}{high T$_c$}
\rn{\Im}{\mathrm{Im}\,}
\nc{\kf}{k_F}
\nc{\kvec}{\vec{k}}
\nc{\lsco}{La$_{2-x}$Sr$_x$CuO$_4$}
\nc{\pecfl}{pECFL}
\nc{\pecmd}{MD-pECFL}
\nc{\pecmi}{MI-pECFL}
\nc{\secfl}{sECFL}
\nc{\tc}{T$_c$}
\nc{\tj}{$t$-$J$}

\begin{document}


\title{A simple phenomenological model for describing the normal state single particle spectral function of high temperature superconductors}


\author{Kazue Matsuyama}
\email{kmatsuya@ucsc.edu}
\affiliation{Department of Physics, University of California, Santa Cruz, CA 95064}
\author{Rohit Dilip}
\affiliation{Irvington High School, Fremont, CA 94538}
\author{G.-H. Gweon}
\email{SGweon@gmail.com}
\affiliation{Department of Physics, University of California, Santa Cruz, CA 95064}


\date{\today}

\begin{abstract}
Describing the normal state single particle spectral function line shapes of high temperature superconductors remains an important goal in condensed matter physics.  Recently\cite{matsuyama_phenomenological_2013}, we have proposed a phenomenological extremely correlated Fermi liquid (pECFL) model that promises to accomplish this goal and that is uniquely distinguished from other models.  Here, we present an even more simplified phenomenological model, which we refer to as the aECFL model, that performs practically at the same level as the pECFL model.  Noting the similarities of the aECFL model and the pECFL model, as well as the differences between the two models, we emphasize the universal significance of the $\omega$-dependence of the so-called caparison factor in the ECFL model.
\end{abstract}

\pacs{Valid PACS appear here}
\maketitle


\section{Introduction}
\label{sec:level1}

In the study of high temperature superconductors, the interpretation of the angle resolved photoelectron spectroscopy (ARPES) data is a major thread of research, which remains vitally important.  Perhaps one of the most challenging, but yet the most basic, subject in this line of research is that of establishing the correct model in the normal state \cite{anderson_strange_2006}.

In our previous work \cite{matsuyama_phenomenological_2013}, we have
introduced a phenomenological model, the phenomenological extremely
correlated electron Fermi-liquid (pECFL) model, that is successful in
explaining both the energy distribution curves (EDCs) and the momentum
distribution curves (MDCs) of ARPES data on equal footing.  The
characteristics of this work can be summarized as follows: (1) High
quality fits of ARPES data are easily obtained for both EDCs and MDCs
within a simple model. (2) The model is based on a Fermi liquid theory,
albeit an unusual one \cite{shastry_extremely_2011,shastry_anatomy_2011},
in contrast to other well-known theoretical models that strive to fit
ARPES data, such as the hidden Fermi liquid (HFL) theory
\cite{anderson_strange_2006} and the marginal Fermi liquid (MFL) theory
\cite{varma_phenomenology_1989}, both of which are non-Fermi liquid
theories. (3) The model is able to explain the universal EDC line
shape (characterized by the strong asymmetry) and the material-dependent
non-universal MDC line shape (which can be asymmetric or symmetric) thanks
to one central feature, the ``caparison factor.''  In contrast, other theories
tend to produce always symmetric MDCs (MFL), or always strongly asymmetric MDCs
and insufficiently asymmetric EDCs (HFL).

Our pECFL model \cite{matsuyama_phenomenological_2013} was the outcome of the evolution of a model \cite{gweon_extremely_2011} that was inspired directly from the original many body theory \cite{shastry_extremely_2011}.  The achievement of the pECFL model is impressive, and the construction of the pECFL model was soundly based on the basic principles such as the positivity of the spectral function and the causality of the Green's function.  Nevertheless, one might wonder whether it is possible to come up with an alternative model, as
the momentum dependent pECFL (MD-pECFL) model described in our previous work \cite{matsuyama_phenomenological_2013} can be viewed as somewhat complicated.

In this work, we show that indeed an alternative model is possible.  We
name this new phenomenological model the aECFL model, where ``a'' stands
for the parameter $a$, the key parameter that governs the behavior of the
model.  Also, one may interpret ``a'' to mean ``alternative.''

As we shall see below, while the pECFL model remains more sound in its construction, the aECFL model has the advantage of being simpler while being able to reproduce all results of the pECFL model.  As a matter of principle, the aECFL model has the dangerous possibility of breaking the positivity of the spectral function, if the value of $a$ becomes too large.  Fortunately, we find that the values of $a$ for both \bittot{} (Bi2212) and \lsco{} (LSCO) remain small.  In this sense, we may regard the aECFL model and the pECFL model as being practically equivalent.

\section{The line shape fits of Bi2212 data}

We shall present the aECFL model fit results for Bi2212 before presenting
the model.  All ECFL model parameters are the same as those used in our
previous work \cite{gweon_extremely_2011,matsuyama_phenomenological_2013}.
For the aECFL model, we have a new parameter $a$, which effectively
replaces the two parameters $a_1$ and $a_2$ in the pECFL model
\cite{matsuyama_phenomenological_2013}.  So, below, we focus on the
parameter $a$.

\subsection{EDC description}

\begin{figure}[t] 
\includegraphics[width=3.3in]{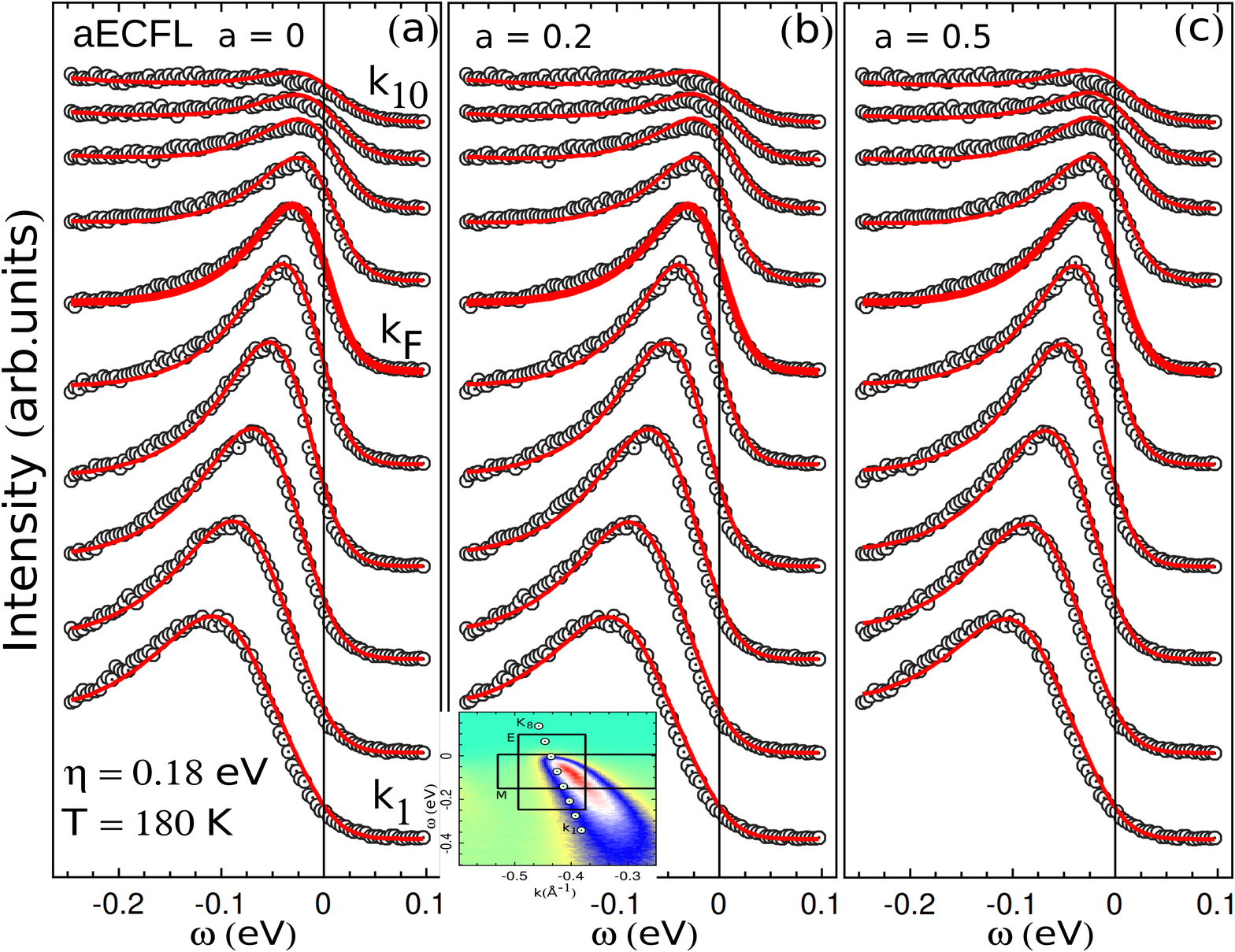}
\caption{Line shape fits of EDCs for Bi2212 ($x$ = 0.15, T = 180 K) (a) aECFL ($a$ = 0) (b) aECFL ($a$ = 0.2) (c) aECFL ($a$ = 0.5) all other fitting parameter values are identical with those used in our previous work \cite{matsuyama_phenomenological_2013}.  The inset in (b) shows the data range used for EDC fits (E) and MDC fits (M; next figure).}
\label{Bi2212EDC}
\end{figure}

Fig.~\ref{Bi2212EDC} shows the EDC curves of Bi2212 taken along the nodal direction fit with the aECFL model.  These data are identical with those of our previous work \cite{matsuyama_phenomenological_2013}.

We see that the value of $a$ does not affect the quality of EDC fits at all.  All fits shown in Fig.~\ref{Bi2212EDC} are equivalent in quality, where the parameter $a$ was taken to be as large as 0.5.

\begin{figure}[b] 
\centerline{\includegraphics[width=3.3in]{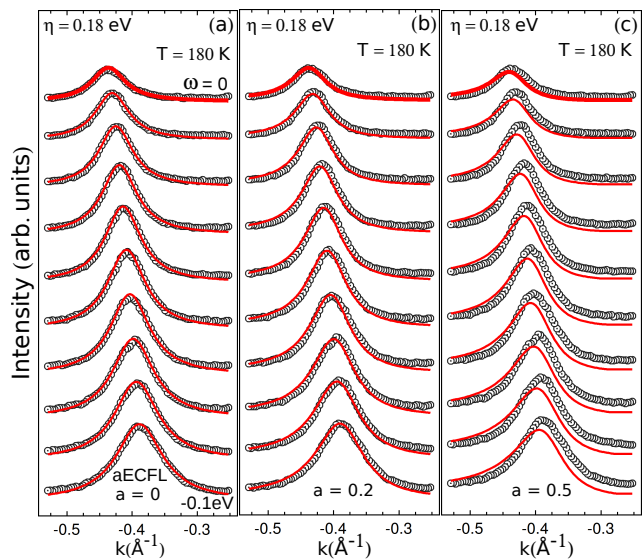}}
\caption{Line shape fits of MDCs for Bi2212 ($x =$ 0.15, T = 180 K) (a) aECFL ($a$ = 0) (b) aECFL ($a$ = 0.2) (c) aECFL ($a$ = 0.5) all other fitting parameter values are identical with those used in our previous work \cite{matsuyama_phenomenological_2013}.}
\label{Bi2212MDC}
\end{figure} 

\subsection{MDC description}

The situation becomes quite different when we examine MDCs for Bi2212.  For MDCs, $a$ makes a crucial difference.  As Fig.~\ref{Bi2212MDC} shows, $a = 0$ gives the best fit.  The fit degrades slightly, but noticeably, when $a = 0.2$.  The degradation is severe and unmistakable when $a = 0.5$.  If $a$ goes past 0.5, the degradation becomes even more severe.

These findings are in good agreement with our previous work \cite{matsuyama_phenomenological_2013}, as we shall see shortly.  At this point, it suffices to note that the symmetry of the MDCs observed for Bi2212 can be described only with $a = 0$.

\section{The $a$ECFL model}
To motivate our new model, let us go back to the original ECFL model, the simple ECFL (sECFL) model \cite{gweon_extremely_2011}:
\begin{align}
	G (\kvec,\omega) &= \frac{Q_n}{\gamma_n} + \frac{\caparison}
		{\omega - \varepsilon(\kvec) - \Phi (\omega)},
		\label{eq-G} \\
	\caparison &= Q_n \left( 1 - \frac{\omega -
			\varepsilon(\kvec)}{\gamma_n} \right).
			& \text{(sECFL)}
		\label{eq-caparison-factor}
\end{align}
where $G$ is the single-particle Green's function and $C_n$ is the so-called ``caparison factor.''  $Q_n = 1 - (n/2)$ where $n$ is the number of electrons per unit cell.  $\gamma_n = 4 Q_n \Delta_0 / n^2$ and $\Delta_0$ ($\sim 0.1$ eV) is an energy scale parameter for $C_n$.  $\varepsilon (\vec k)$ is the one-electron band dispersion relation\footnote{Throughout this paper, we use $\hbar = 1$, and so, e.g., $\omega$ and $\varepsilon (\vec k)$ have the same dimension.}.  $\Phi$ is the Dyson self energy of the underlying Fermi liquid, which was termed the auxiliary Fermi liquid, whose full Green's function is given by $1/(\omega - \varepsilon (\vec k) - \Phi (\omega))$.

In our previous work \cite{matsuyama_phenomenological_2013}, our goal was to develop a systematic way to deal with the problem that the above caparison factor, if used for $\omega \gg \varepsilon (\vec k)$ (as for an MDC analysis), becomes a negative value, leading to an unphysical negative value for the spectral function\footnote{As in our previous work, we work with the advanced Green's function.} $A (\vec k, \omega) = \frac{1}{\pi}\Im G$.

The aECFL idea arises from a simpler idea for remedy to this problem.  In using the above theory to fit ARPES data, the problem lies with the $\varepsilon (\vec k)$ factor, since $\omega$ is limited to mostly negative values for ARPES data.  So, what if we just scale $\varepsilon (\vec k) \rightarrow a \varepsilon (\vec k)$ with $0 \le a \le 1$?

This way, we obtain the caparison factor for the aECFL
\begin{align}
		\caparison &= Q_n \left( 1 - \frac{\omega -
			a\varepsilon(\kvec)}{\gamma_n} \right).
			& \text{(aECFL)}
		\label{eq-caparison-factor-aECFL}
\end{align}
In comparison to our previous work \cite{matsuyama_phenomenological_2013}, this caparison factor is much simpler, but we cannot assure the same rigorous analytical properties (the sum rule and the non-negativity) of the spectral function as we were able to do with our pECFL model \cite{matsuyama_phenomenological_2013}.  Here, our goal is rather to stay with the original spirit of the sECFL model, i.e., the construction of a low energy model that can describe well the data near the Fermi surface.  The strength of the aECFL model is that it can describe MDCs well (as we show here), which the original sECFL could not do.  Also, it is tempting to make a guess that $a$ could arise through some renormalization process \footnote{Private communication, R. R. P. Singh.} although it remains to be seen whether a microscopic calculation will in fact produce such a scaling factor.

Note that, when $a = 0$, we recover the momentum-independent (MI) pECFL model, used in our previous work \cite{matsuyama_phenomenological_2013}.  If $a = 1$, then we recover the sECFL model.

Therefore, it is not surprising at all that the $a= 0$ aECFL model does an excellent job describing EDCs and MDCs for Bi2212, since we have already shown that the MI-pECFL model provides a good description for the Bi2212 data \cite{matsuyama_phenomenological_2013}.

As we will show in the next section, it is possible to quantify this finding further using the aECFL model.  Also, we are able to fit the LSCO data with the same aECFL model \footnote{In contrast, note that in our previous work, the MD-pECFL model, used to fit the LSCO data, and the MI-pECFL model, used to fit the Bi2212 data, were rather different.  The MD-pECFL model interpolates between the sECFL model and a scaled auxiliary FL model, and it does not converge to the MI-pECFL model for any parameter values.}  Here are the main results: $a=0\pm0.03 $ for Bi2212 and $a=0.2\pm0.08$ for LSCO.

\begin{figure}[b] 
\centerline{\includegraphics[width=3.3in]{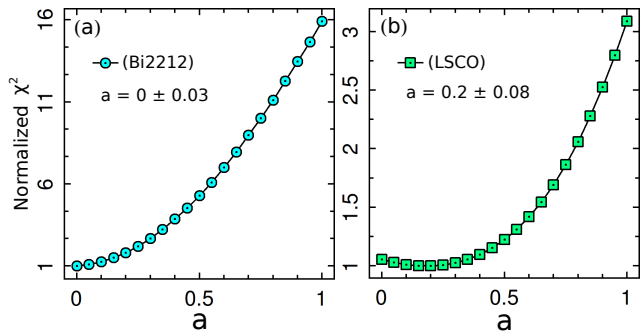}}
\caption{$\chi^2$ variation as a function of $a$.  For each value of $a$, the $\chi^2$ value averaged over all EDCs and MDCs are plotted.  (a) Bi2212 ($x =$ 0.15, T = 180 K). (b) LSCO ($x = $ 0.15, T = 20 K).}
\label{chi2plot}
\end{figure} 
\begin{figure}[b] 
\centerline{\includegraphics[width=3.3in]{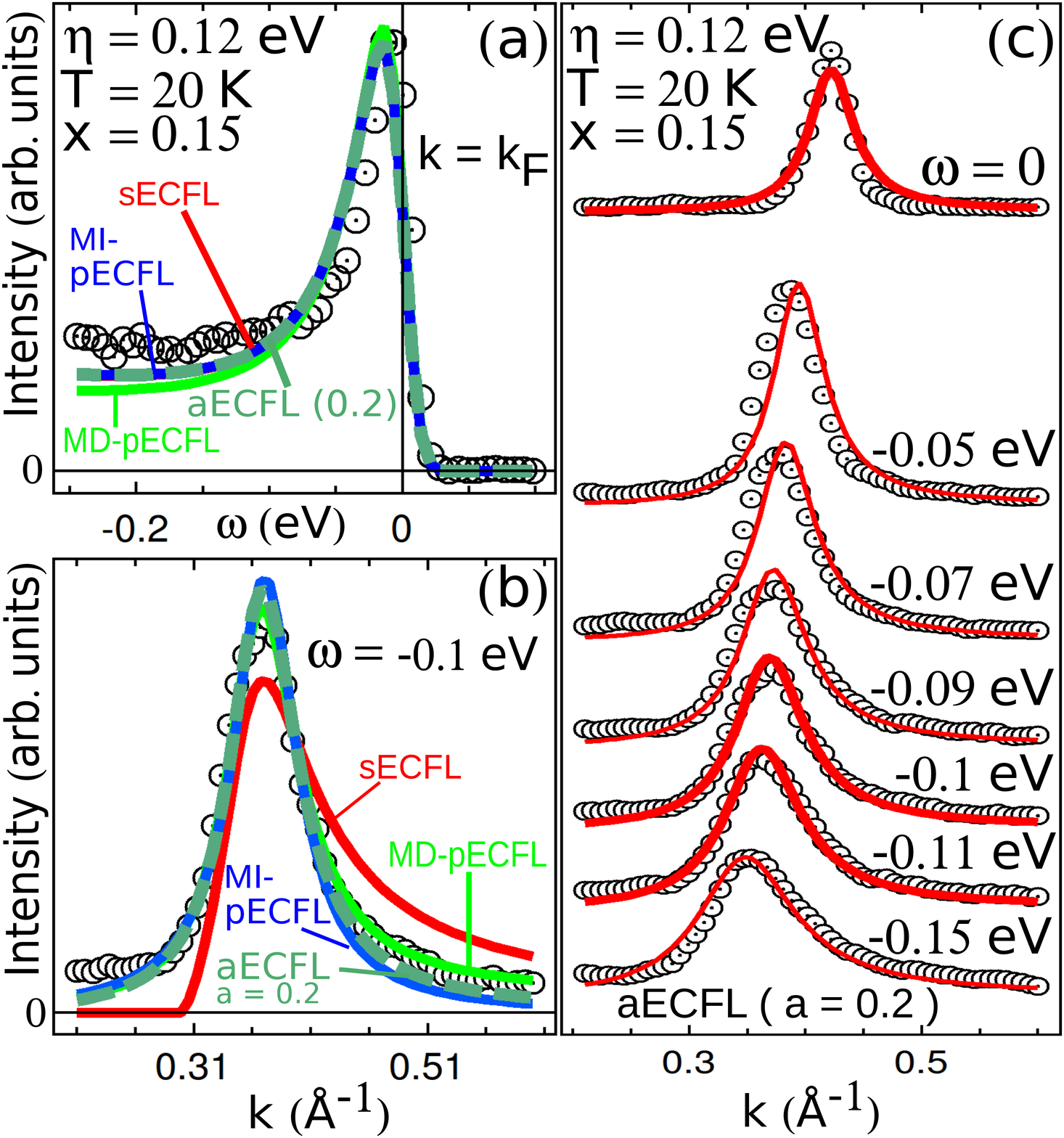}}
\caption{aECFL line shape fits ($a = 0.2$) to the data taken along the nodal direction of optimally doped LSCO \cite{yoshida_low-energy_2007} ($x = 0.15$, T= 20 K) and their comparison with other ECFL models. (a) EDC fits at k = $k_F$, (b) MDC fits at $\omega = -0.1$ eV, (c) MDC line shape fits for $\omega$ values ranging from 0 to $-0.15$ eV.  For the exact range of the fit data, see the next figure (Fig.~\ref{chi2LSCO}(d)).}
\label{LSCOfits}
\end{figure} 

\section{\label{sec:level 2} determination of $a$ value}

Fig.~\ref{chi2plot} shows the values of $\chi^2$ plotted as a function of the parameter $a$ for Bi2212 (a) and LSCO (b).  The displayed $\chi^2$ value is the average of the average $\chi^2$ value of EDC fits and the average $\chi^2$ value of MDC fits. We determined the value of $\chi^2$ in the data range used for EDC fits, and we will describe this reason in the next section when we present the line shape fits of LSCO data.  As evidenced by these graphs, the qualities of fits degrade significantly as the parameter $a$ becomes large.

To determine the best value of the parameter $a$ for each material shown in Fig.~\ref{chi2plot}, we read the $a$ value for the minimum position of $\chi^2$. We find $a = 0 $ for Bi2212 and $a = 0.2$ for LSCO\@.  As Fig.~\ref{chi2plot} shows, the minimum position of $\chi^2$ is well-defined, giving only a small uncertainty, $\pm 0.03$, for the estimated $a$ value.  We can estimate the uncertainty of $a$ further by adjusting the fit ranges of EDCs and MDCs and then examining the new values of $a$.  The adjustment of the fitting ranges by $\pm 20\%$ resulted in little change to the $\chi^2$ plot of Bi2212, while the minimum position of $\chi^2$ of LSCO shifted by $\pm 0.08$.  Therefore, the uncertainty for $a$ value is $\pm 0.03$ for Bi2212 and $\pm 0.08$ for LSCO\@.

As noted already, regardless of $a$ values, all ECFL model parameter values
were held fixed as those same values used in our previous work
\cite{matsuyama_phenomenological_2013}.  It is also worth noting that the
bare Fermi velocity is treated as a model-dependent, i.e., $a$-dependent,
quantity as in the same work \cite{matsuyama_phenomenological_2013}. Its value
(times $\hbar$, which is set to 1 in our work) for Bi2212 was varied linearly
from 6.3 eV\AA{} ($a = 0$; MI-pECFL) to 5.5 eV\AA{} ($a = 1$; sECFL) and that
for LSCO were varied linearly from 5.5 eV\AA{} ($a = 0$; MI-pECFL) to 4 eV\AA{}
($a = 1$; pECFL).

\section{\label{sec:level 2} The line shape fits of LSCO data}

Next, we present the fits of LSCO data by the aECFL model in
Fig.~\ref{LSCOfits}.  Fig.~\ref{LSCOfits}(a) shows the EDC fitting by the aECFL model at the Fermi momentum.  Fitting quality comparison between all phenomenological ECFL models show no difference, and fits are all equally successful.  Fig.~\ref{LSCOfits}(b) shows the MDC fitting at $\omega = - 0.1 $ eV.  Here, we see that the fit quality varies over different ECFL models.   The aECFL fit is very good, almost as good as the best fit (the MD-pECFL fit).  A more quantitative discussion of the fit quality will be presented below.

In Fig.~\ref{LSCOfits}(c), we show MDC fits by the aECFL model at various energy values.  The aECFL model successfully describes the asymmetry in MDC curves which becomes more obvious for higher energy values ($-\omega \gtrsim 80$ meV).  As the theory is for the normal state, and the data are for the superconducting state, the comparison between theory and experiment is not straightforward.  As we argued in our previous work \cite{matsuyama_phenomenological_2013}, however, the comparison is still quite meaningful at high energy values, where the measured electronic structure remains essentially unchanged in the normal state.  Even at low energy, the comparison between theory and experiment remains quite good.

The description of the LSCO data with the aECFL model is as good as the
description with the pECFL model in our previous work
\cite{matsuyama_phenomenological_2013}.  To examine the quality of the aECFL fit a bit more in depth, we now examine the
$\chi^2$ values for different ECFL models.

\begin{figure}[t] 
\centerline{\includegraphics[width=3.3in]{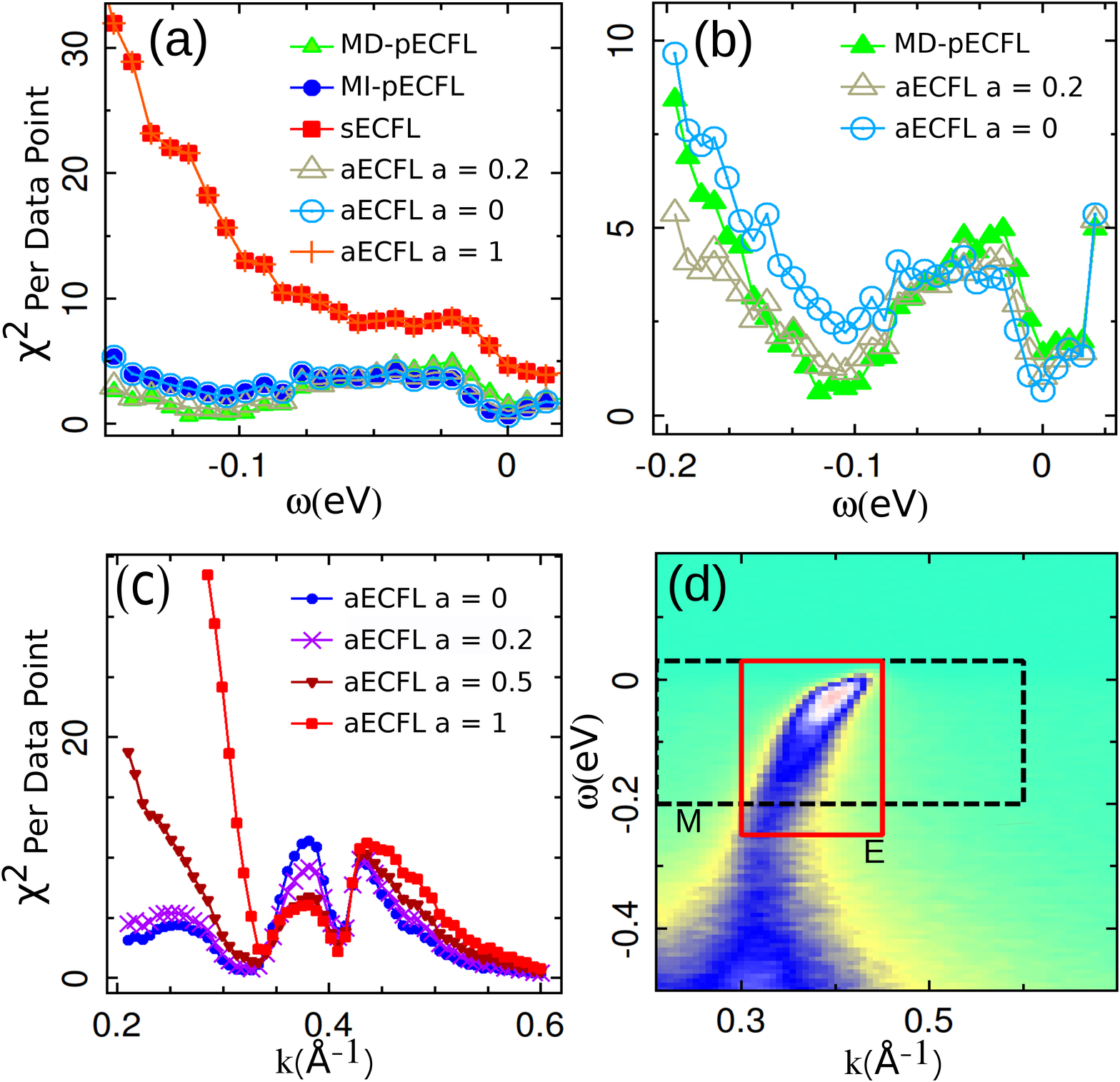}}
\caption{(a) Comparison of $\chi^2$ values for MDC fits using various ECFL models.  (b) The same plot as (a) but for only three ECFL models, showing details.  (c) $\chi^2$ values for EDC fits by aECFL. (d) LSCO data along the nodal direction \cite{yoshida_low-energy_2007} ($x = 0.15 $, T = 20 K). Two rectangles show the data ranges used for the MDC fits (M) and the EDC fits (E), to produce analysis results reported in this paper.  However, one exception is panel (c) of this figure, where EDC fits are shown for a much wider ranges of $k$.}
\label{chi2LSCO}
\end{figure} 

In Fig.~\ref{chi2LSCO}, we show $\chi^2$ value per data point (a,b,c), as well as the data ranges used in our fits (d).
In Fig.~\ref{chi2LSCO}(a), we see that the aECFL models with the value of $a$ from 0 to 1 are able to reproduce the same good MDC fits as pECFL fits.  On a closer look, we can discern differences, as we can do with  Fig.~\ref{chi2LSCO}(b).  In this figure, we can see that on average the aECFL model with $a = 0.2$  performs the best, when we focus on the important high energy range ($-\omega \gtrsim 80 $ meV).  When $a$ increases, the fit quality degrades: this is shown dramatically for the $a = 1$ case (for which aECFL becomes sECFL) in panel (a).

When we look at the $\chi^2$ values for EDC fits, the same pathology for large $a$ can be noticed.  In panel (c), we show $\chi^2$ values for EDC fits.  For $k$ values corresponding to deep inside the Fermi surface ($k \lesssim 0.3$ \AA$^{-1}$), we see that the fit quality degrades dramatically when $a$ increases beyond about 0.5.  The origin of this pathology is, of course, the sign problem of the caparison factor of the sECFL model, as we discussed above.

Within the aECFL formalism presented here, therefore, we limited the momentum range used for EDC fits to that indicated in Fig.~\ref{chi2LSCO}(d) to avoid the deep-inside-Fermi-surface region.  We also emphasize that a large value of $a$ ($\gtrsim 0.5$) would make the aECFL model ill-defined, and so care must be taken that the application of this model be limited to small $a$ values.

\section{\label{sec:level 2} Conclusion}

In this work, we showed that it is possible to construct an alternative model, called aECFL, to our previous phenomenological ECFL models. The aECFL model is simple and is as successful as our previously reported models.

The dichotomy of EDC line shape and MDC line shape has been explained well by our previous and current models, and other theory has been proposed  \cite{ovchinnikov_general_2014} to explain the same.  The particularly nice aspect of the current aECFL model is that (1) the asymmetry of EDC is guaranteed by the strong $\omega$ dependence of the caparison factor, while (2) the degree of the asymmetry of MDC is tuned by the $a$ parameter.


\bibliography{aECFL}

\begin{thebibliography}{12}%
\makeatletter
\providecommand \@ifxundefined [1]{%
 \@ifx{#1\undefined}
}%
\providecommand \@ifnum [1]{%
 \ifnum #1\expandafter \@firstoftwo
 \else \expandafter \@secondoftwo
 \fi
}%
\providecommand \@ifx [1]{%
 \ifx #1\expandafter \@firstoftwo
 \else \expandafter \@secondoftwo
 \fi
}%
\providecommand \natexlab [1]{#1}%
\providecommand \enquote  [1]{``#1''}%
\providecommand \bibnamefont  [1]{#1}%
\providecommand \bibfnamefont [1]{#1}%
\providecommand \citenamefont [1]{#1}%
\providecommand \href@noop [0]{\@secondoftwo}%
\providecommand \href [0]{\begingroup \@sanitize@url \@href}%
\providecommand \@href[1]{\@@startlink{#1}\@@href}%
\providecommand \@@href[1]{\endgroup#1\@@endlink}%
\providecommand \@sanitize@url [0]{\catcode `\\12\catcode `\$12\catcode
  `\&12\catcode `\#12\catcode `\^12\catcode `\_12\catcode `\%12\relax}%
\providecommand \@@startlink[1]{}%
\providecommand \@@endlink[0]{}%
\providecommand \url  [0]{\begingroup\@sanitize@url \@url }%
\providecommand \@url [1]{\endgroup\@href {#1}{\urlprefix }}%
\providecommand \urlprefix  [0]{URL }%
\providecommand \Eprint [0]{\href }%
\providecommand \doibase [0]{http://dx.doi.org/}%
\providecommand \selectlanguage [0]{\@gobble}%
\providecommand \bibinfo  [0]{\@secondoftwo}%
\providecommand \bibfield  [0]{\@secondoftwo}%
\providecommand \translation [1]{[#1]}%
\providecommand \BibitemOpen [0]{}%
\providecommand \bibitemStop [0]{}%
\providecommand \bibitemNoStop [0]{.\EOS\space}%
\providecommand \EOS [0]{\spacefactor3000\relax}%
\providecommand \BibitemShut  [1]{\csname bibitem#1\endcsname}%
\let\auto@bib@innerbib\@empty
\bibitem [{\citenamefont {Matsuyama}\ and\ \citenamefont
  {Gweon}(2013)}]{matsuyama_phenomenological_2013}%
  \BibitemOpen
  \bibfield  {author} {\bibinfo {author} {\bibfnamefont {K.}~\bibnamefont
  {Matsuyama}}\ and\ \bibinfo {author} {\bibfnamefont {G.-H.}\ \bibnamefont
  {Gweon}},\ }\href {\doibase 10.1103/PhysRevLett.111.246401} {\bibfield
  {journal} {\bibinfo  {journal} {Physical Review Letters}\ }\textbf {\bibinfo
  {volume} {111}},\ \bibinfo {pages} {246401} (\bibinfo {year}
  {2013})}\BibitemShut {NoStop}%
\bibitem [{\citenamefont {Anderson}(2006)}]{anderson_strange_2006}%
  \BibitemOpen
  \bibfield  {author} {\bibinfo {author} {\bibfnamefont {P.~W.}\ \bibnamefont
  {Anderson}},\ }\href {\doibase 10.1038/nphys388} {\bibfield  {journal}
  {\bibinfo  {journal} {Nat Phys}\ }\textbf {\bibinfo {volume} {2}},\ \bibinfo
  {pages} {626} (\bibinfo {year} {2006})}\BibitemShut {NoStop}%
\bibitem [{\citenamefont
  {Shastry}(2011{\natexlab{a}})}]{shastry_extremely_2011}%
  \BibitemOpen
  \bibfield  {author} {\bibinfo {author} {\bibfnamefont {B.~S.}\ \bibnamefont
  {Shastry}},\ }\href {\doibase 10.1103/PhysRevLett.107.056403} {\bibfield
  {journal} {\bibinfo  {journal} {Physical Review Letters}\ }\textbf {\bibinfo
  {volume} {107}},\ \bibinfo {pages} {056403} (\bibinfo {year}
  {2011}{\natexlab{a}})}\BibitemShut {NoStop}%
\bibitem [{\citenamefont {Shastry}(2011{\natexlab{b}})}]{shastry_anatomy_2011}%
  \BibitemOpen
  \bibfield  {author} {\bibinfo {author} {\bibfnamefont {B.~S.}\ \bibnamefont
  {Shastry}},\ }\href {\doibase 10.1103/PhysRevB.84.165112} {\bibfield
  {journal} {\bibinfo  {journal} {Physical Review B}\ }\textbf {\bibinfo
  {volume} {84}},\ \bibinfo {pages} {165112} (\bibinfo {year}
  {2011}{\natexlab{b}})}\BibitemShut {NoStop}%
\bibitem [{\citenamefont {Varma}\ \emph {et~al.}(1989)\citenamefont {Varma},
  \citenamefont {Littlewood}, \citenamefont {Schmitt-Rink}, \citenamefont
  {Abrahams},\ and\ \citenamefont {Ruckenstein}}]{varma_phenomenology_1989}%
  \BibitemOpen
  \bibfield  {author} {\bibinfo {author} {\bibfnamefont {C.~M.}\ \bibnamefont
  {Varma}}, \bibinfo {author} {\bibfnamefont {P.~B.}\ \bibnamefont
  {Littlewood}}, \bibinfo {author} {\bibfnamefont {S.}~\bibnamefont
  {Schmitt-Rink}}, \bibinfo {author} {\bibfnamefont {E.}~\bibnamefont
  {Abrahams}}, \ and\ \bibinfo {author} {\bibfnamefont {A.~E.}\ \bibnamefont
  {Ruckenstein}},\ }\href {\doibase 10.1103/PhysRevLett.63.1996} {\bibfield
  {journal} {\bibinfo  {journal} {Physical Review Letters}\ }\textbf {\bibinfo
  {volume} {63}},\ \bibinfo {pages} {1996} (\bibinfo {year}
  {1989})}\BibitemShut {NoStop}%
\bibitem [{\citenamefont {Gweon}\ \emph {et~al.}(2011)\citenamefont {Gweon},
  \citenamefont {Shastry},\ and\ \citenamefont {Gu}}]{gweon_extremely_2011}%
  \BibitemOpen
  \bibfield  {author} {\bibinfo {author} {\bibfnamefont {G.-H.}\ \bibnamefont
  {Gweon}}, \bibinfo {author} {\bibfnamefont {B.~S.}\ \bibnamefont {Shastry}},
  \ and\ \bibinfo {author} {\bibfnamefont {G.~D.}\ \bibnamefont {Gu}},\ }\href
  {\doibase 10.1103/PhysRevLett.107.056404} {\bibfield  {journal} {\bibinfo
  {journal} {Physical Review Letters}\ }\textbf {\bibinfo {volume} {107}},\
  \bibinfo {pages} {056404} (\bibinfo {year} {2011})}\BibitemShut {NoStop}%
\bibitem [{Note1()}]{Note1}%
  \BibitemOpen
  \bibinfo {note} {Throughout this paper, we use $\hbar = 1$, and so, e.g.,
  $\omega $ and $\varepsilon (\protect \mathaccentV {vec}17Ek)$ have the same
  dimension.}\BibitemShut {Stop}%
\bibitem [{Note2()}]{Note2}%
  \BibitemOpen
  \bibinfo {note} {As in our previous work, we work with the advanced Green's
  function.}\BibitemShut {Stop}%
\bibitem [{Note3()}]{Note3}%
  \BibitemOpen
  \bibinfo {note} {Private communication, R. R. P. Singh.}\BibitemShut {Stop}%
\bibitem [{Note4()}]{Note4}%
  \BibitemOpen
  \bibinfo {note} {In contrast, note that in our previous work, the MD-pECFL
  model, used to fit the LSCO data, and the MI-pECFL model, used to fit the
  Bi2212 data, were rather different. The MD-pECFL model interpolates between
  the sECFL model and a scaled auxiliary FL model, and it does not converge to
  the MI-pECFL model for any parameter values.}\BibitemShut {Stop}%
\bibitem [{\citenamefont {Yoshida}\ \emph {et~al.}(2007)\citenamefont
  {Yoshida}, \citenamefont {Zhou}, \citenamefont {Lu}, \citenamefont {Komiya},
  \citenamefont {Ando}, \citenamefont {Eisaki}, \citenamefont {Kakeshita},
  \citenamefont {Uchida}, \citenamefont {Hussain}, \citenamefont {Shen},\ and\
  \citenamefont {Fujimori}}]{yoshida_low-energy_2007}%
  \BibitemOpen
  \bibfield  {author} {\bibinfo {author} {\bibfnamefont {T.}~\bibnamefont
  {Yoshida}}, \bibinfo {author} {\bibfnamefont {X.~J.}\ \bibnamefont {Zhou}},
  \bibinfo {author} {\bibfnamefont {D.~H.}\ \bibnamefont {Lu}}, \bibinfo
  {author} {\bibfnamefont {S.}~\bibnamefont {Komiya}}, \bibinfo {author}
  {\bibfnamefont {Y.}~\bibnamefont {Ando}}, \bibinfo {author} {\bibfnamefont
  {H.}~\bibnamefont {Eisaki}}, \bibinfo {author} {\bibfnamefont
  {T.}~\bibnamefont {Kakeshita}}, \bibinfo {author} {\bibfnamefont
  {S.}~\bibnamefont {Uchida}}, \bibinfo {author} {\bibfnamefont
  {Z.}~\bibnamefont {Hussain}}, \bibinfo {author} {\bibfnamefont {Z.-X.}\
  \bibnamefont {Shen}}, \ and\ \bibinfo {author} {\bibfnamefont
  {A.}~\bibnamefont {Fujimori}},\ }\href {\doibase
  10.1088/0953-8984/19/12/125209} {\bibfield  {journal} {\bibinfo  {journal}
  {Journal of Physics: Condensed Matter}\ }\textbf {\bibinfo {volume} {19}},\
  \bibinfo {pages} {125209} (\bibinfo {year} {2007})}\BibitemShut {NoStop}%
\bibitem [{\citenamefont {Ovchinnikov}\ \emph {et~al.}(2014)\citenamefont
  {Ovchinnikov}, \citenamefont {Shneyder},\ and\ \citenamefont
  {Kordyuk}}]{ovchinnikov_general_2014}%
  \BibitemOpen
  \bibfield  {author} {\bibinfo {author} {\bibfnamefont {S.~G.}\ \bibnamefont
  {Ovchinnikov}}, \bibinfo {author} {\bibfnamefont {E.~I.}\ \bibnamefont
  {Shneyder}}, \ and\ \bibinfo {author} {\bibfnamefont {A.~A.}\ \bibnamefont
  {Kordyuk}},\ }\href {\doibase 10.1103/PhysRevB.90.220505} {\bibfield
  {journal} {\bibinfo  {journal} {Phys. Rev. B}\ }\textbf {\bibinfo {volume}
  {90}},\ \bibinfo {pages} {220505} (\bibinfo {year} {2014})}\BibitemShut
  {NoStop}%
\end{thebibliography}%
\end{document}